# PRODUCTION OF NEUTRON-RICH HYPERNUCLEI AT DAFNE2

Vincenzo Paticchio *

for the FINUDA Collaboration

*INFN Sezione di Bari, via Amendola, 173, 70126 Bari, Italy

*Abstract*

The search for neutron-rich Hypernuclei is connected to many aspects of hypernuclear physics and of astrophysics. The discovery of these exotic Hypernuclei is possible by the Finuda experiment coupled to a DaΦne upgrade to a luminosity of $L = 10^{34}$ cm$^{-2}$s$^{-1}$, a condition at which their production rates are here evaluated considering the possible mechanisms.

## Physics Motivations

The first motivation at all to search for neutron-rich Hypernuclei is the confirmation of their existence. The discovery of a neutron-rich Hypernucleus is, from several years, a challenge for experimental nuclear physicists.

In fact, in spite of many theoretical works [1,2,3] which predict their production, almost no experimental data confirm the existence of neutron-rich Λ Hypernuclei and of the related production mechanism.

Until now, only poor experimental data have been produced in a series of Λ and Σ hypernuclear studies at KEK using the ($K^-_{stopped}$, $\pi^+$) reaction [4] while, more recently, in an experiment at the K6 beam line of KEK 12-GeV PS, the ($\pi^-$,$K^+$) double charge-exchange reaction has been measured to produce $^{10}_\Lambda$Li, $^{12}_\Lambda$Be, the results of which measurements are not available yet.

Further motivations arise from one of the most exciting recent discoveries in nuclear physics, i.e. the observation of the neutron halo phenomenon (low density matter).

Hypernuclei can be even better candidates than ordinary nuclei to study nuclear matter with extreme N/Z ratios (i.e.$^7_\Lambda$H, $^6_\Lambda$H, $^{12}_\Lambda$Be), because more extended mass distributions are expected than in ordinary nuclei thanks to the glue-like role of the Λ , and its effect on neutron-halo (see fig.1) [6, 7].

In the field of astrophysics a deeper knowledge of the ΛN –ΣN interaction is welcome because the presence of hyperons in high density nuclear matter is expected to explain various phenomena in neutron stars and helps to describe better their Equation of State [8, 9, 10].

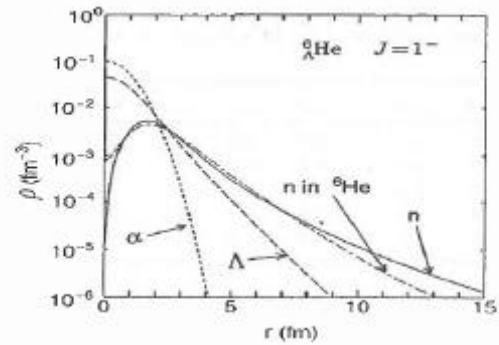

Figure 1: The mass distributions of $^6_\Lambda$He with the Λ in the halo is expected to be more extended than that $^6$He [7]. The glue-like role of the Λ explains because the $^6_\Lambda$He is bound while $^5$He is unbound .

## Production Mechanisms

Typical neutron-rich Λ hypernuclei like $^{12}_\Lambda$Be, $^{16}_\Lambda$C, and $^{11}_\Lambda$Li can be produced by means of the ($K^-$, $\pi^+$) reaction with $K^-$ at rest.

A first mechanism for this reaction is a two-step process with strangeness and charge exchange :

$$K^- + p \rightarrow \Lambda + \pi^0, \quad \pi^0 + p \rightarrow n + \pi^+ \quad . \qquad (1)$$

Another possible mechanism is the production of Λ hypernuclei in one-step process via an admixture of a virtual Σ$^-$ state due the Λ–Σ coupling :

$$K^- + p \rightarrow \Sigma^- + \pi^+, \quad \Sigma^- + p \leftrightarrow n + \Lambda \quad . \qquad (2)$$

Recently theoretical calculations were performed [2], in which the production rates relative to several light nuclear targets were calculated for the ($K^-$, $\pi^+$) reactions both in single step and in two steps. The $^{12}_\Lambda$Be expected production rates (P$_{HYP}$) for the two mechanisms are in Table 1.

Table1: Production rates expected for the reaction $^{12}C(K,\pi^+)^{12}_\Lambda Be$ [2].

| $J^\pi$ | $P_{HYP}$ (1) | $P_{HYP}$ (2) |
|---|---|---|
| $1^-$ | $1.8 \times 10^{-5}$ | $1.2 \times 10^{-6}$ |
| $0^+$ | $6.0 \times 10^{-6}$ | $1.6 \times 10^{-7}$ |

## Yield Estimation and MonteCarlo Simulation

The Finuda expected counting rates ($R_{HIP}$) are obtained by

$$R_{HIP} = \sigma_\Phi \cdot L \cdot \Delta t \cdot BR(K^-K^+) \cdot P_{HYP} \cdot \varepsilon_{det} \quad (3)$$

where:

- $\sigma_\Phi$ : $\Phi$ at rest production cross section ($\sigma_\Phi = 3,26$ [11]);
- $L$ : DaΦne luminosity;
- $\Delta t$ : data taking duration;
- $P_{HYP}$ : neutron-rich hypernuclei production rates;
- $\varepsilon_{det}$ : Finuda global efficiency ($\varepsilon_{det} = 0.082$ [12]).

Then for the $^{12}_\Lambda Be$, using the production rates of Table1, at a DaΦne luminosity of $L = 10^{34}$ cm$^{-2}$s$^{-1}$ we obtain:

$$R_{HIP} = 85 \text{ counts/h}.$$

Now, we consider the production of the very exotic neutron-rich hypernucleus $^6_\Lambda H$, characterised by a large neutron excess (N/Z = 4). Following the same hypothesis used by Saha et al.[5] for the production rates of $^{10}_\Lambda Li$, we can expect for $^6_\Lambda H$ a production rate one order of magnitude larger than the one of $^{12}_\Lambda Be$.

For the production of $^{12}_\Lambda Be$, in the process of charge exchange a nucleon is forced to change drastically its state, because the quantum numbers of the external neutron always differ from the ones of the protons. Such a drastic change is not needed for the $^6_\Lambda H$ production. For this reason the production of $^6_\Lambda H$ is enhanced.

Then, substituting in (3) a reasonable $P_{HYP} = 1.0 \times 10^{-4}$ we can expect for $^6_\Lambda H$:

$$R_{HIP} = 474 \text{ counts/h}.$$

Additional reactions, different from (1) and (2) of above, are able to produce a $\pi^+$ with a momentum larger than 200 MeV/c. In particular the in-flight decay of a $\Sigma^+$ from:

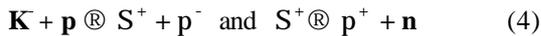

$$K^- + p \to \Sigma^+ + \pi^- \text{ and } \Sigma^+ \to \pi^+ + n \quad (4)$$

The reactions (1) and (2) produce the $\pi^+$ peaks relative to hypernuclei formation while the last one produces the most important source of continuous background. As can be seen from above, the $\pi^+$ background events are associated in the reaction (4) to the production of a 100÷180 MeV/c $\pi^-$, which can be detected and rejected by the Finuda spectrometer in about half of the cases.

The Finuda MonteCarlo simulation was used, for both cases here considered, to evaluate the signal to background ratio and the amount of time needed to obtain hypernuclear peaks clearly out of the statistical fluctuation of the background (see table 2 and fig.2).

Table 2: Hypernuclear Peaks Identification (if $S > 2\sigma_B$)

| $^Z_\Lambda A$ | $J^\pi$ | S/B | Time |
|---|---|---|---|
| $^{12}_\Lambda Be$ | $1^-$ | 0.05 | 37 m |
| $^{12}_\Lambda Be$ | $0^+$ | 0.01 | 8h 14 m |
| $^6_\Lambda H$ | $0^+$ | 0.24 | 90 s |
| $^6_\Lambda H$ | $1^-$ | 0.02 | 2h 45m |

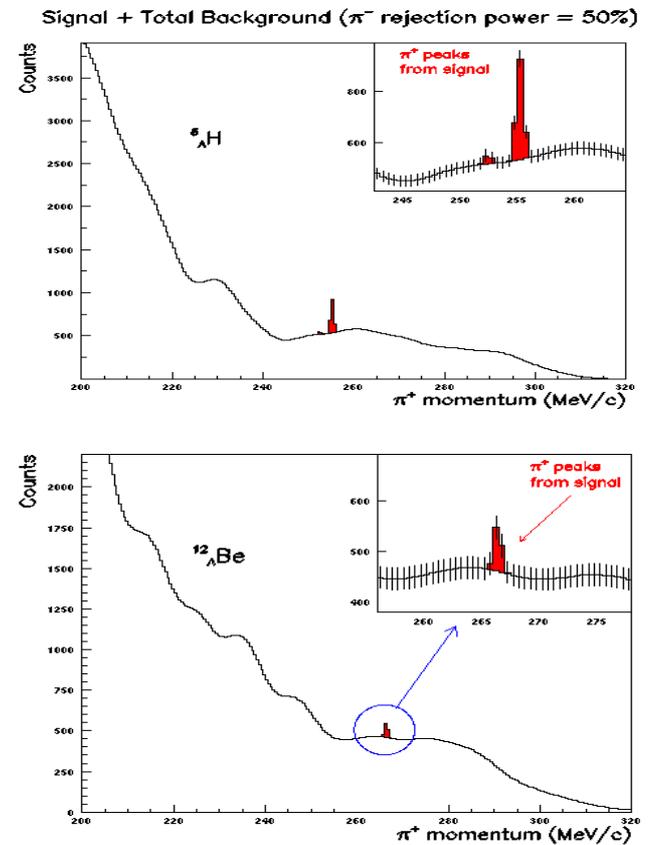

Figure 2: Finuda Montecarlo simulated $\pi^+$ spectra for $^6_\Lambda H$ and $^{12}_\Lambda Be$ formation. In the latter case, the states $1^-$ e $0^+$ are not resolved.

## Conclusions

The Finuda spectrometer and DaΦne2 upgraded in luminosity ($L = 10^{34}$) allow to perform a high quality neutron-rich hypernuclei identification with very large statistics and hence the possibility to study also their structure and decays.